\documentclass[preprint,review, 12pt,sort, compress]{elsarticle}

\usepackage{amssymb}
\usepackage{graphicx}
\usepackage{dcolumn}
\usepackage{bm}
\usepackage[utf8]{inputenc}
\usepackage[T1]{fontenc}
\usepackage{mathptmx}
\usepackage{etoolbox}
\usepackage{url}
\usepackage{hyperref}
\usepackage{amsmath,amsthm, amssymb} 
\usepackage{graphicx}
\usepackage{enumerate} 
\usepackage{float, subfig} 
\usepackage{color} 
\usepackage{xspace} 
\usepackage{array} 
\usepackage{braket} 
\usepackage{ulem}
\usepackage{relsize}
\usepackage{tabularx}
\usepackage[nodisplayskipstretch]{setspace}
\usepackage{gensymb} 
\usepackage{bm} 
\usepackage{array}
\newcolumntype{P}[1]{>{\centering\arraybackslash}p{#1}}

\renewcommand{\bold}[1]{\boldsymbol{#1}} 

\renewcommand{\v}[1]{\ensuremath{\mathbf{#1}}} 
\newcommand{\gv}[1]{\ensuremath{\mbox{\boldmath$ #1 $}}}
\newcommand{\avg}[1]{\left< #1 \right>} 
\newcommand{\pd}[2]{\frac{\partial #1}{\partial #2}} 

\newcommand{\mode}{\!\left(\substack{\gv{\kappa} \\ \nu}\right)}

\newcommand {\eqn}[1] {Eq.~(\ref{#1})}

\newcommand {\fig}[1] {Fig.~\ref{#1}}

\newcommand {\figs}[1] {Figs.~\ref{#1}}
\newcommand {\figx}[1] {\ref{#1}}

\newcommand{\deltmd}{\Delta t_{\rm MD}}

\journal{Carbon}

\begin{document}

\begin{frontmatter}



\title{Lattice thermal conductivity and phonon properties of polycrystalline graphene}

\author{Kunwar Abhikeern}
\author{Amit Singh\corref{cor1}}\ead{amit.k.singh@iitb.ac.in}


\affiliation{organization={Department of Mechanical Engineering, IIT Bombay},
            addressline={}, 
            city={Mumbai},
            postcode={40076}, 
            state={Maharashtra},
            country={India}}

\begin{abstract}
Using spectral energy density method, we predict the phonon scattering mean lifetimes of polycrystalline graphene (PC-G) having polycrystallinity 
only along $\rm{x}$-axis with seven different misorientation (tilt) angles at room
temperature. Contrary to other studies on PC-G samples, our results indicate 
strong dependence of the thermal conductivity (TC) on the tilt angles. We also show 
that the square of the group velocity components along $\rm{x}$ and $\rm{y}$ axes and the phonon
lifetimes are uncorrelated and the phonon density of states are almost the same for 
all samples with different tilt angles. Further, a  distribution of the group 
velocity component along $\rm{x}$ or $\rm{y}$ axis as function of normal frequency is found to be exponentially decaying whereas that of phonon lifetime showed piecewise 
constant function behavior with respect to frequency. We provide parameters for these distribution functions and suggest another measure of the TC based on these distributions. Finally, we perform a size-dependent analysis for two tilt angles, $21.78^\circ$ and $32.20^\circ$, and find that bulk TC components decrease by around 34\% to 62\% in comparison to the bulk TC values of the pristine graphene. Our analysis reveals intriguing insights into the interplay between grain orientation, phonon scattering and thermal conductivity in graphene. 
\end{abstract}

\begin{graphicalabstract}
\includegraphics[width=1.0\textwidth]{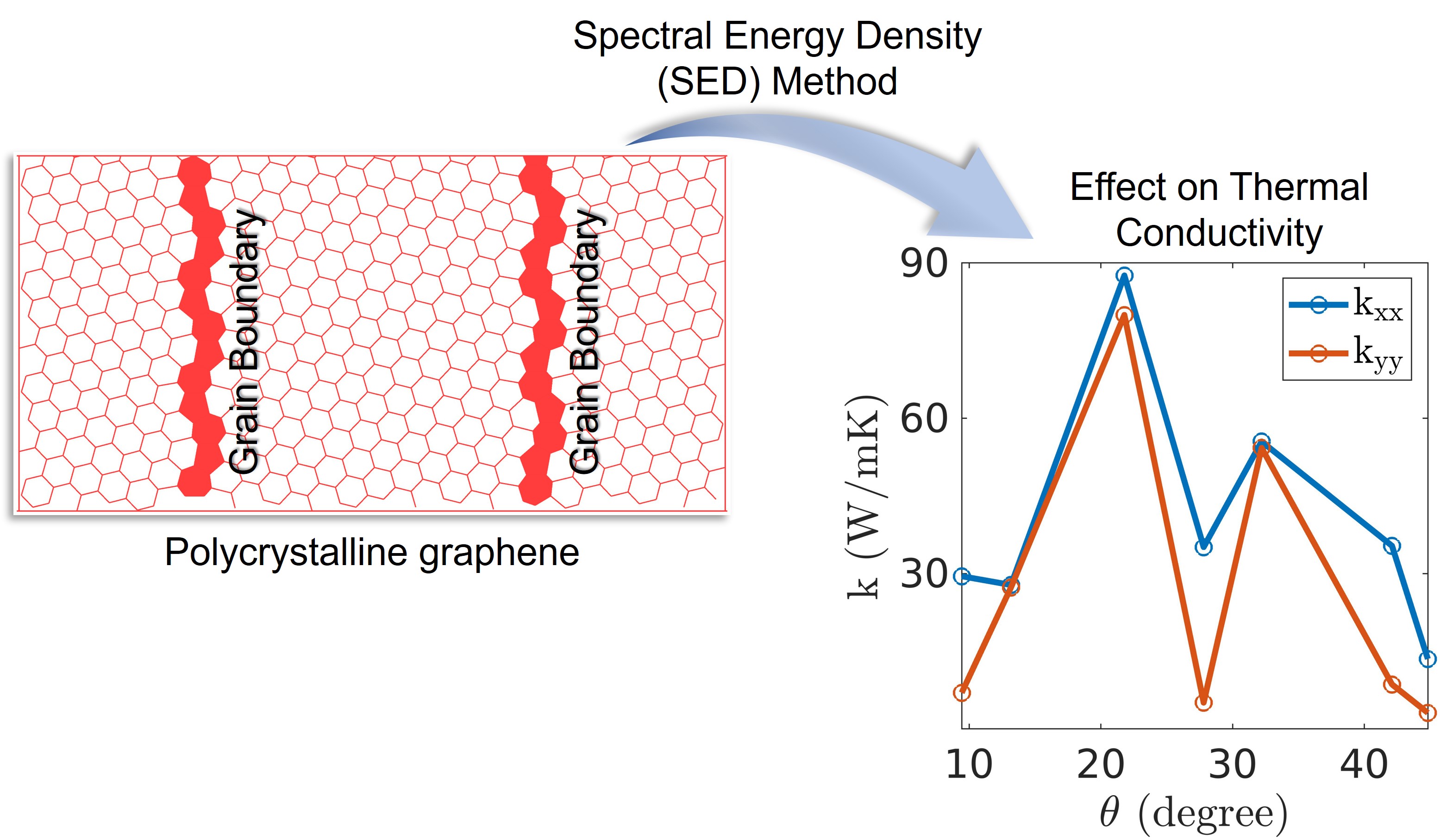}
\end{graphicalabstract}

\begin{highlights}
\item TC of polycrystalline graphene strongly depends upon misorientation angles of the GBs.
\item Distribution of the group velocity components against frequency exponentially decays.
\item Distribution of phonon lifetime shows piecewise constant function behaviour.
\item A measure of the thermal conductivity based on these distributions has been suggested.
\end{highlights}

\begin{keyword}


Polycrystalline graphene \sep Grain boundary \sep Lattice thermal conductivity \sep Spectral energy density method \sep Phonon properties
\end{keyword}

\end{frontmatter}


\section{Introduction}\label{sec:Intro}
The computational and experimental thermal conductivity (TC) of single and bilayer graphene systems has received significant attention over the years ~\cite{abhikeern2023consistent}. 
One crucial aspect often overlooked is the influence of grain boundaries (GBs) on the TC of polycrystalline graphene (PC-G) with some notable exceptions. 

Experimentally, it has been found that the TC of PC-G increases with increase in grain size.
Ma et al~\cite{ma2017} reported that the TC exponentially increased from almost 610 to 5230~W/mK at the room temperature as the grain size increased from almost 200~nm to 10~$\mu$m. 
Similarly, Woomin Lee et al~\cite{lee2017} found the in-plane TC in the temperature range from 
320 K to 510 K to be 680--340, 1890--1020 and 2660--1230~W/mK for average grain sizes of 
0.5, 2.2, and 4.1~$\mu$m, respectively. Dongmok Lee et al~\cite{lee2017dependence} measured
significantly
smaller TCs (412-572~W/mK) for PC-G samples with reduced grain sizes of less than 1~$\mu$m.
They also showed almost no dependence on misorienation (tilt) angles of GBs for the TCs of PC-G
samples whereas there was strong dependence on misorienation angles for bicrystalline graphene (BC-G) samples.

Numerically, one of the earliest studies to investigate the GB effect on the TCs of PC-G samples
was done by Bagri et al \cite{bagri2011thermal}, who performed reverse nonequilibrium molecular dynamics simulations with modified Tersoff interatomic potential on tilt GBs based graphene samples with misorientation angles $5.5^\circ$, 
$13.2^\circ$ and $21.7^\circ$, and estimated the bulk TCs to be around 2220, 2380 and 2380~W/mK, respectively, thereby showing a very week dependence on the misorientation angles.
They also calculated boundary conductance (inverse of Kapitza resistance) which decreased with increase in misorientation angles. 
Following nonequilibrium Green’s function approach, Serov et al~\cite{serov2013effect} 
showed strong increasing dependence of the TCs of PC-G samples  on grain sizes ranging from  close to 1 nm to 1000~nm at room temperature, for which 
the TCs vary from  close to 100~W/mK to 550~W/mK. However, the dependence on the misorientation angle was found to be negligible.
In another study, using REBO potential for interatomic potential and Green-Kubo method for
the calculation of the TC, Mortazavi et al~\cite{mortazavi2014} prepared PC-G samples 
with average grain size from 1~nm to 5~nm and found the TCs of PC-G samples to be one order of magnitude smaller than that of pristine graphene. With the same REBO potential and approach-to-equilibrium molecular dynamics simulations, Hahn et al~\cite{hahn2016thermal} estimated the bulk TC of PC-G with an average grain size of 1 nm as 26.6 W/mK at 300~K, which was almost 3\% of the crystalline sample. The presence of GBs significantly decreased the estimated mean free path from 451~nm for crystalline graphene to around 30~nm for PC-G, which suggested increased scattering of phonons with GBs. With the help of Green-Kubo method and optimized Tersoff potential, Liu et al~\cite{liu2014grain}
showed that the TC decreases exponentially with increasing GB energy for PC-G samples.  

With the help of the nonequilibrium molcular dynamics (NEMD) method~\cite{singh:2015a}, another set of computational works \cite{cao2012kapitza,zhang2012thermal,azizi2017kapitza}  on BC-G samples  show strong dependence of the tilt angle and average density defects along GB on the Kapitza resistance.
In a detailed comprehensive NEMD based study, Fox et al~\cite{fox2019thermal} performed simulations on bicrystalline graphene nanoribbons (bi-GNR) \footnote{Although they write polycrystalline but their simulation details suggest bi-GNR samples because periodicity along x-axis is not present.} for 
a wide range of tilt angles ($\theta$) under arbitrary in-plane thermal loading directions ($\phi)$. They
showed that the TCs decrease from 0$^\circ$ to 32.2$^\circ$ tilt angles and then increase 
almost symmetrically up to 60$^\circ$ tilt angles. Having done the 
size-dependent analysis for $\phi=10^\circ$, they calculated the bulk TCs for $\theta = 9.4^\circ, 32.2^\circ, 44.8^\circ$ and found them to be close to 416, 312 and 476~W/m K, respectively. This dependence on misorientation angles is similar to the experimental work
done by Dongmok Lee et al~\cite{lee2017dependence} for BC-G samples.
Using similar NEMD method, another study found that 10.98 $\degree$ BC-G displays anomalous higher TC compared to other misorientaion angles \cite{liu2014anomalous}, which was not 
observed by Fox et al~\cite{fox2019thermal}. In both works, it has been established that the TC is inversely proportional to the dislocation density of GB. 
In a recent NEMD based work done by Tong et al~\cite{tong2021phononic}, where the heat flux was calculated with the incorrect thermostat work done method~\cite{abhikeern2023consistent}, which
overestimates the TC value in comparison to that obtained by a more consistent 
Irving-Kirkwood based calculation of heat flux, the TCs of BC-G samples were found to first decrease with the tile angle from
0$^\circ$ to 13.18$^\circ$, but then contrary to Fox et al~\cite{fox2019thermal} was shown to have increased for 21.78$^\circ$ tilt angle.

These works mostly focus upon the calculation of the TCs of PC-G or BC-G samples using 
Green-Kubo or NEMD approaches~\cite{singh:2015a,singh:2015b} and show that the TCs increase
with increasing grain size. The TC of BC-G depends upon the misorientation angle~\cite{lee2017dependence,  liu2014anomalous, fox2019thermal,liu2014grain}, however, the
dependence has been found to be weak for PC-G samples~\cite{bagri2011thermal,serov2013effect}.
The NEMD based works also explore the effect of GBs and misorientation angles on the
Kapitza conductance and some studies have also been able to capture the dependence on GB energy and dislocation density, however, none of these works investigate the microstructural properties
of the TC of PC-G samples such as phonon mean lifetimes and group velocities of different phonon modes in detail. We found only one study done by Tong et al~\cite{tong2021phononic} which 
discusses the contribution of phonon lifetimes and group velocities, obtained from the Spectral Energy Density (SED) method and the lattice dynamics, respectively, for the GBs with three different misorientation angles, $\theta = 4.41^\circ, 13.18^\circ, 21.78^\circ$, however, the treatment and the discussion remain brief. Moreover, the sample preparation does not seem to have followed proper procedures to maintain stable dislocation core structure~\cite{blase2000structure}. 
The study also adopts the alternative phonon SED $\Phi^{\prime}$~\cite{larkin2014comparison}
approach for the extraction of spectral phonon relaxation times which does not involve eigenvectors and, therefore, does not represent the  phonon spectral energy~\cite{larkin2014comparison} accurately. Further, the calculation of phonon lifetimes and group velocities has been performed for polycrystalline samples because the SED method assumes periodicity along all three coordinate axes, however, they have been cited as supporting evidences for the TCs of B-CG samples obtained by the NEMD method.

In the present work, we study the thermal transport in polycrystalline graphene samples with polycrystallinity only along $\rm{x}$-axis (xPC-G samples) with the help of the robust and consistent 
normal mode decomposition (NMD) analysis based SED method~\cite{abhikeern2023consistent,mcgaughey2014predicting}. 
These samples cannot be categorized under BC-G samples or randomly oriented PC-G samples in the references cited above.
First, almost similar size samples with seven different symmetric misorientation angles and stable dislocation cores are prepared and then the phonon properties and the 
TC tensor components, $\rm{k}_{\rm{xx}}$ and $\rm{k}_{\rm{yy}}$, are calculated for each sample. A pristine graphene sample of the same size is also studied for comparison. 
 This is followed by a size-dependent analysis of thermal transport for
pristine graphene and xPC-G samples with two different misorientation angles, 
$21.78^\circ$ and $32.2^\circ$, 
so that bulk TCs of graphene with GBs of different misorientations can be compared.
Contrary to other works on PC-G samples cited above, our work establishes strong 
dependence of the TCs of xPC-G samples on misorientation angles. 
 
The structure of the paper is as follows. Section 2 describes the sample preparation method for 
xPC-G samples with different misorientation angles. In Section 3, we perform MD 
simulations necessary for the SED analysis. We discuss the results in Section 4 
after calculating the phonon
group velocities with the help of the harmonic lattice dynamics based GULP package and 
phonon lifetimes by fitting the SED curves to the Lorentzian function.
In this section, we also obtain bulk TCs of three graphene systems with the SED method. 
Finally, we conclude in Section 5 with a summary and suggestions for any future work.

\section{Sample preparation}
\begin{figure*}
    \centering
    \subfloat[SLG]{\includegraphics[width=0.285\textwidth]{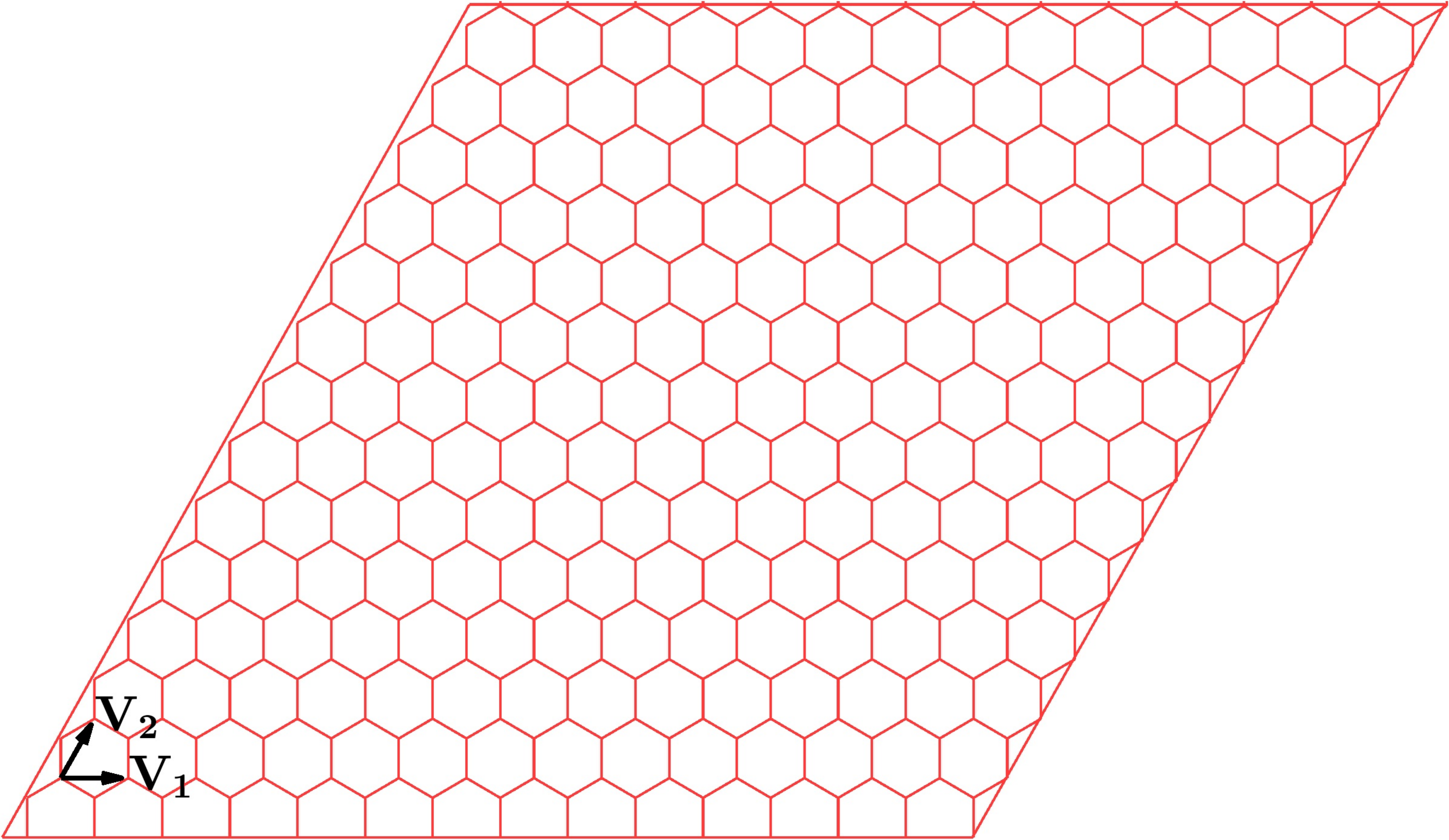}\label{fig01a}}
    \hspace{0.1cm}
    \subfloat[(3,4)~9.43\degree]{\includegraphics[width=0.33\textwidth]{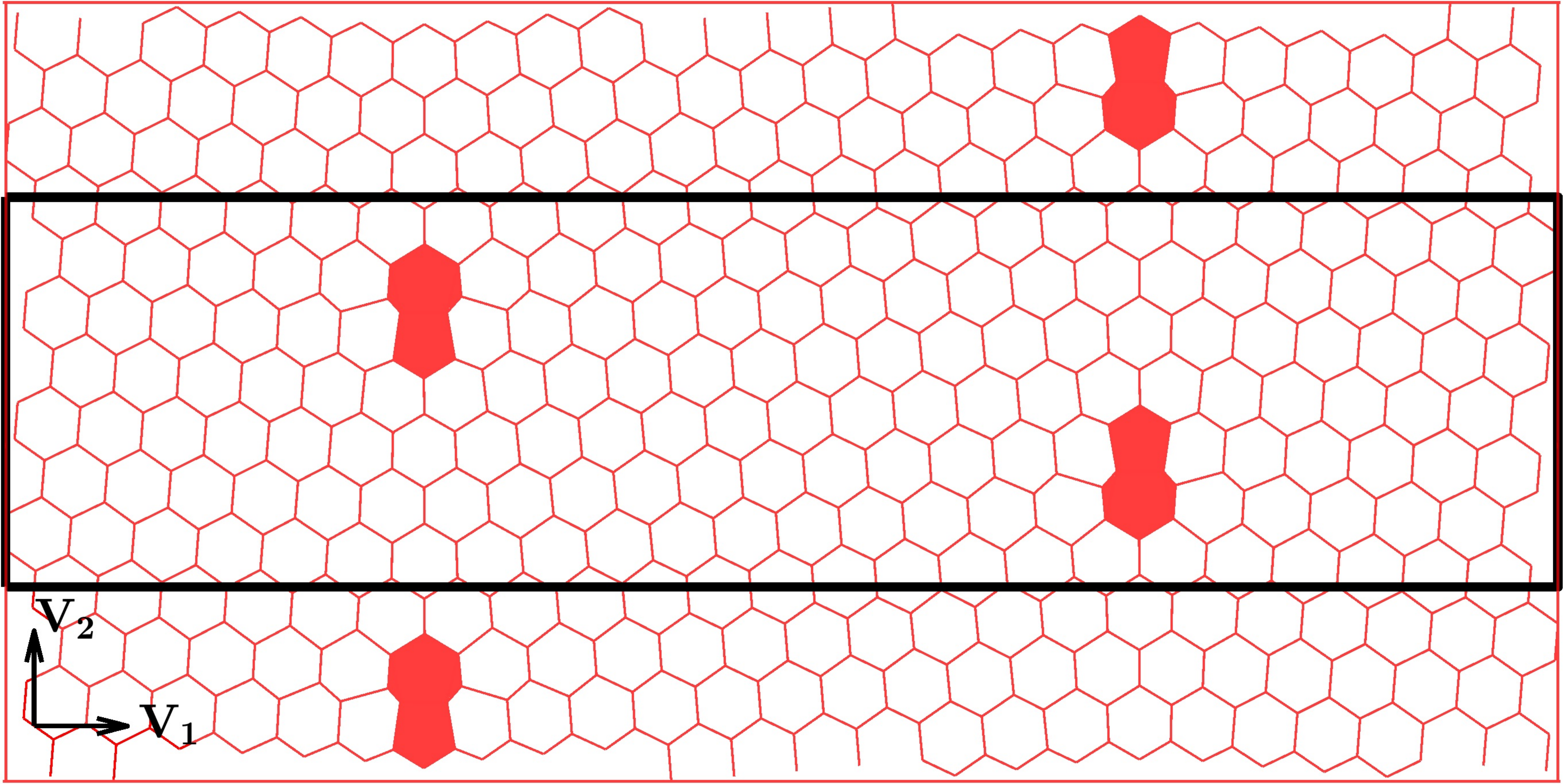}\label{fig01b}}
    \hspace{0.1cm}
    \subfloat[(2,3)~13.17\degree ]{\includegraphics[width=0.33\textwidth]{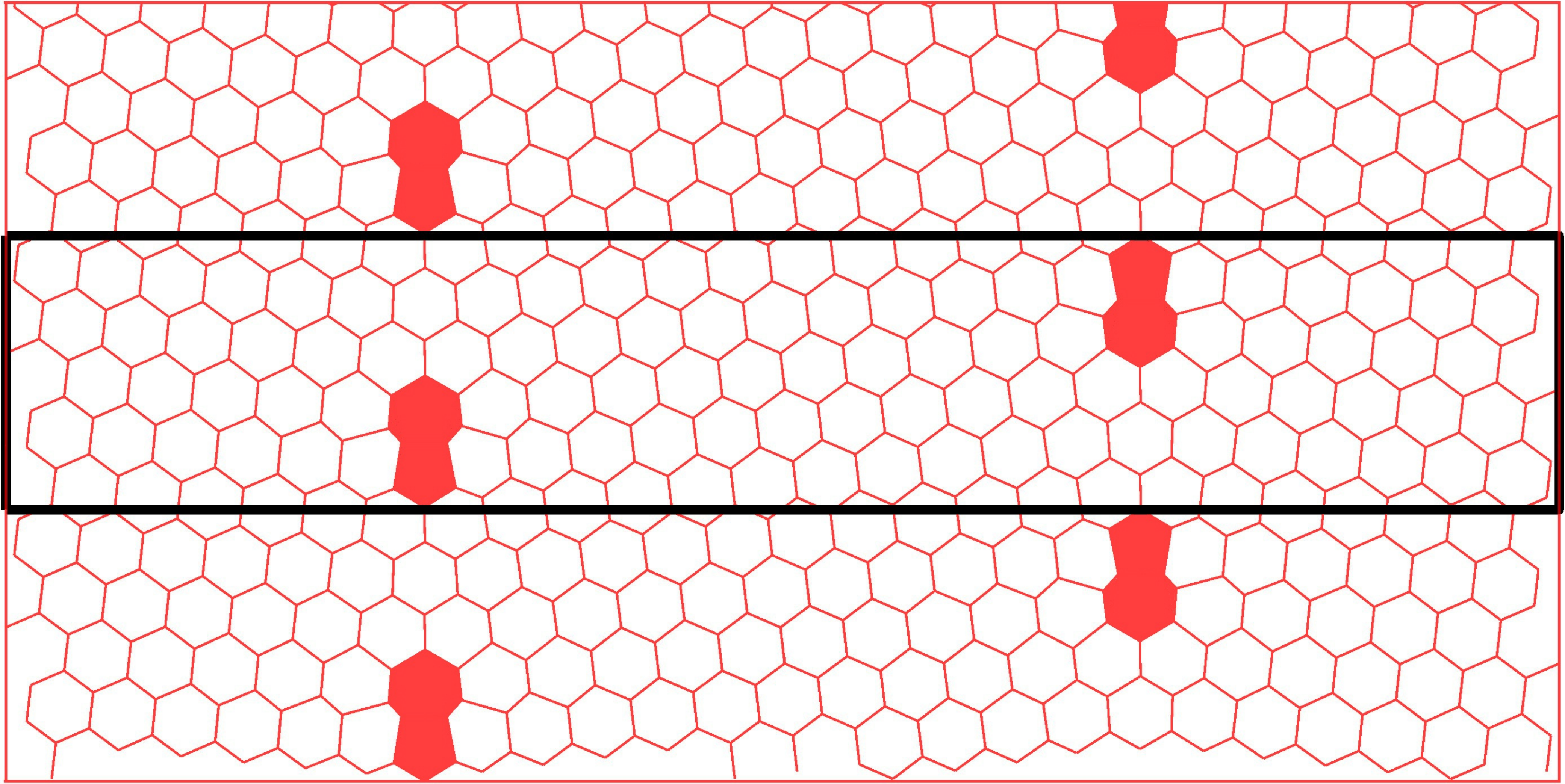}\label{fig01c}}\\
    \hspace{0.3cm}
    \subfloat[(1,2)~21.78\degree ]{\includegraphics[width=0.33\textwidth]{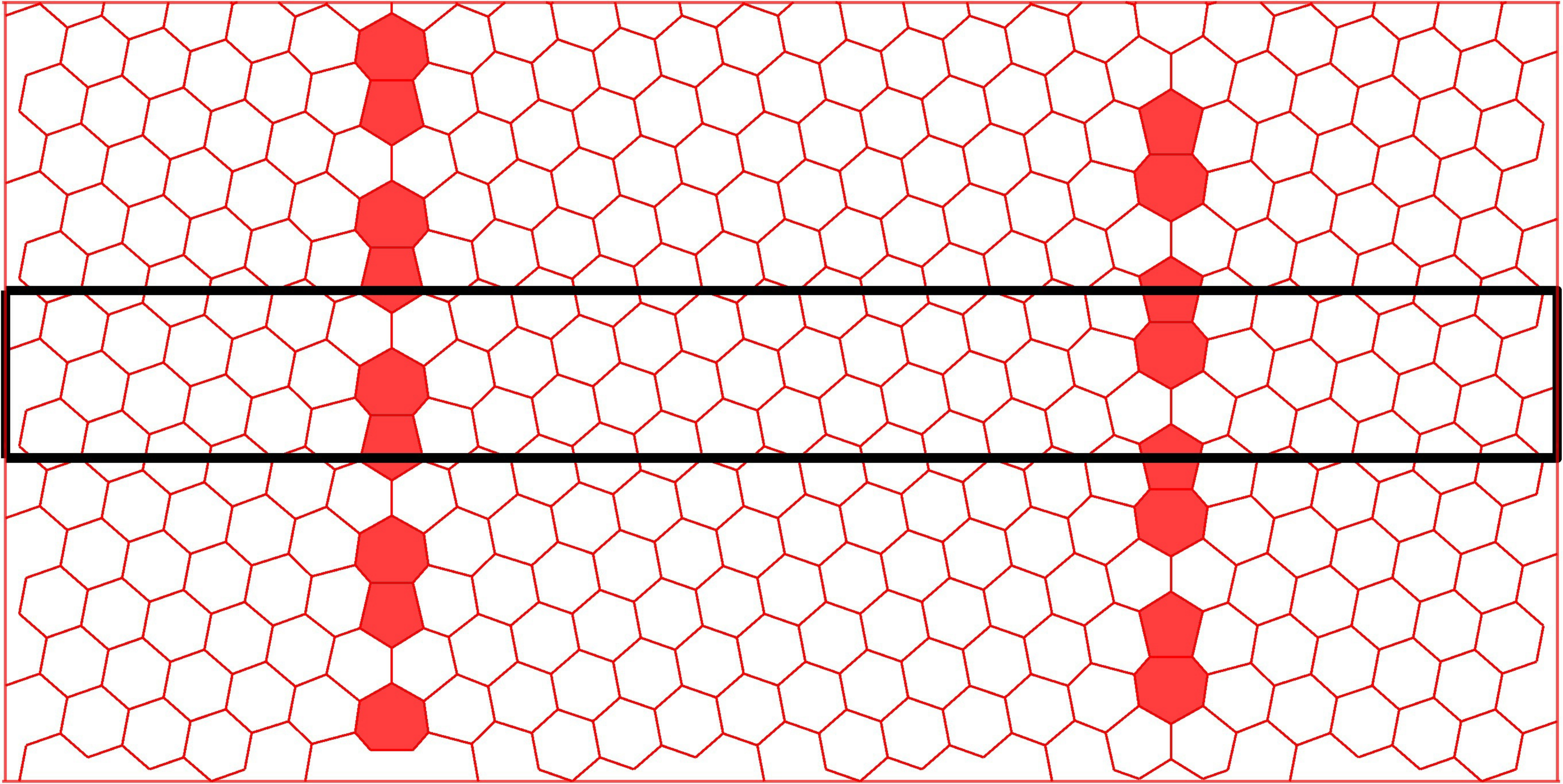}\label{fig01d}}
    \hspace{0.1cm}
    \subfloat[(2,5) ~27.80\degree ]{\includegraphics[width=0.33\textwidth]{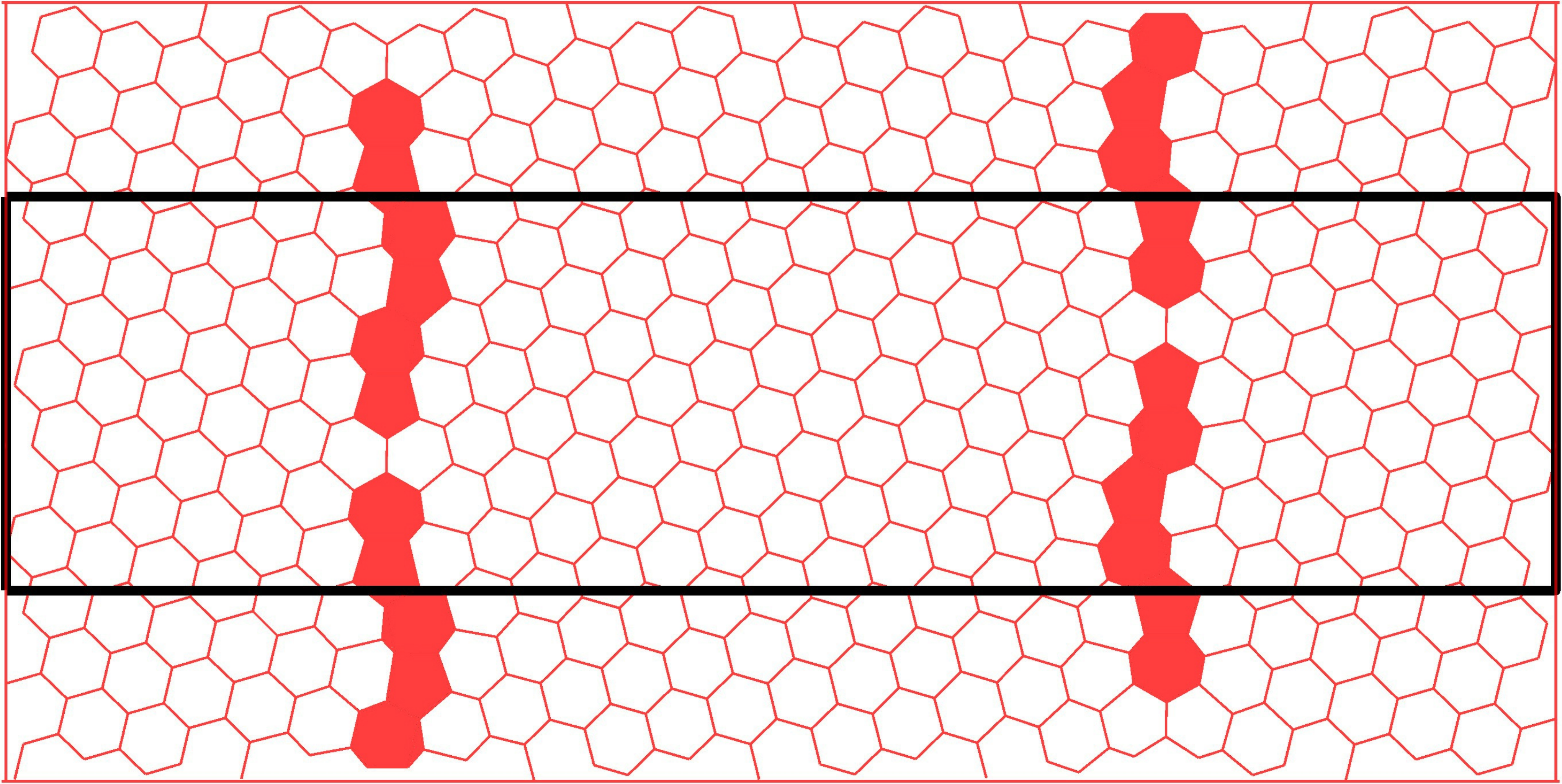}\label{fig01e}}\\
    \subfloat[(1,3) ~32.20\degree]{\includegraphics[width=0.32\textwidth]{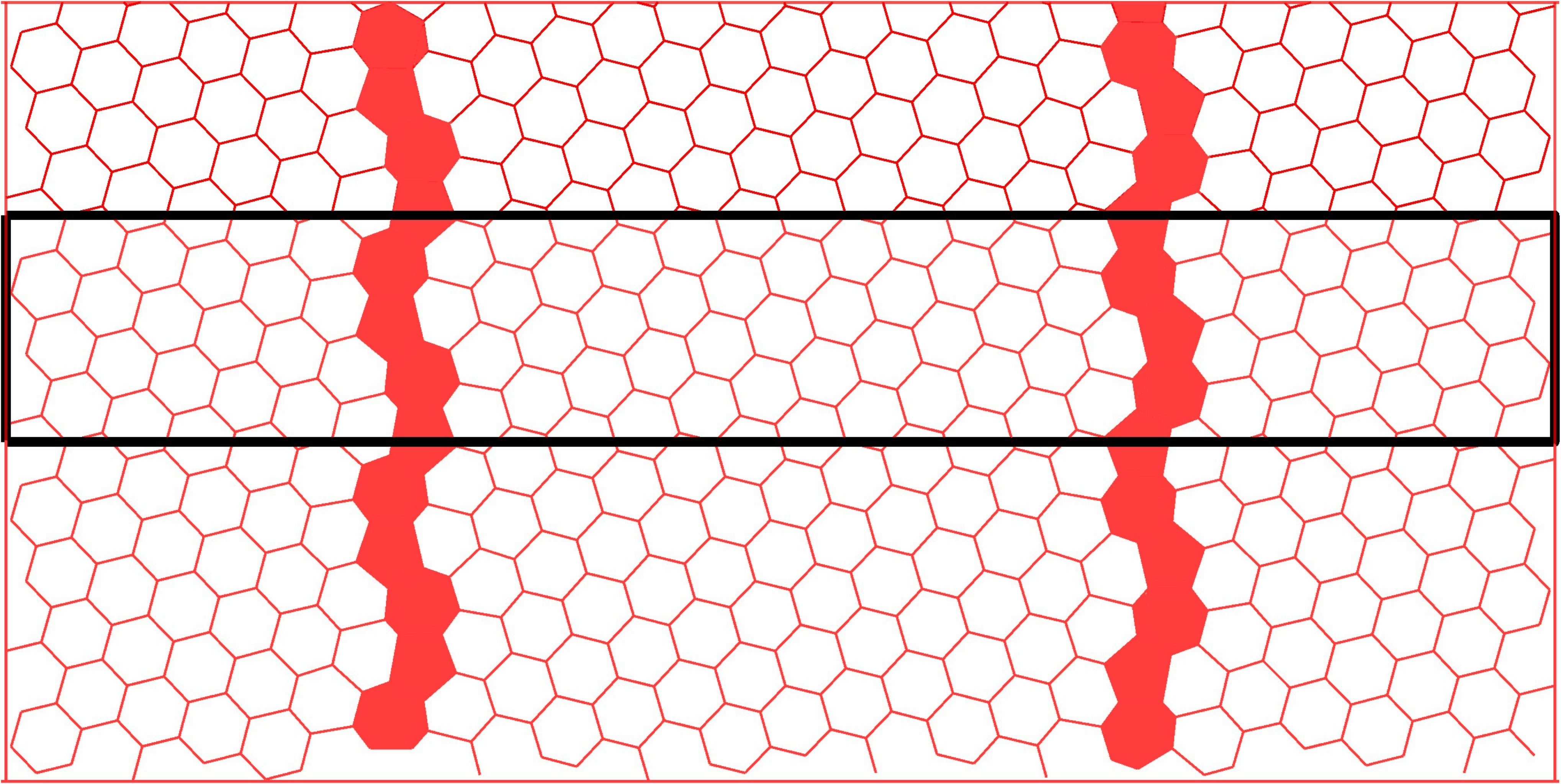}\label{fig01f}}
    \hspace{0.1cm}
    \subfloat[(1,5)~42.10\degree ]{\includegraphics[width=0.32\textwidth]{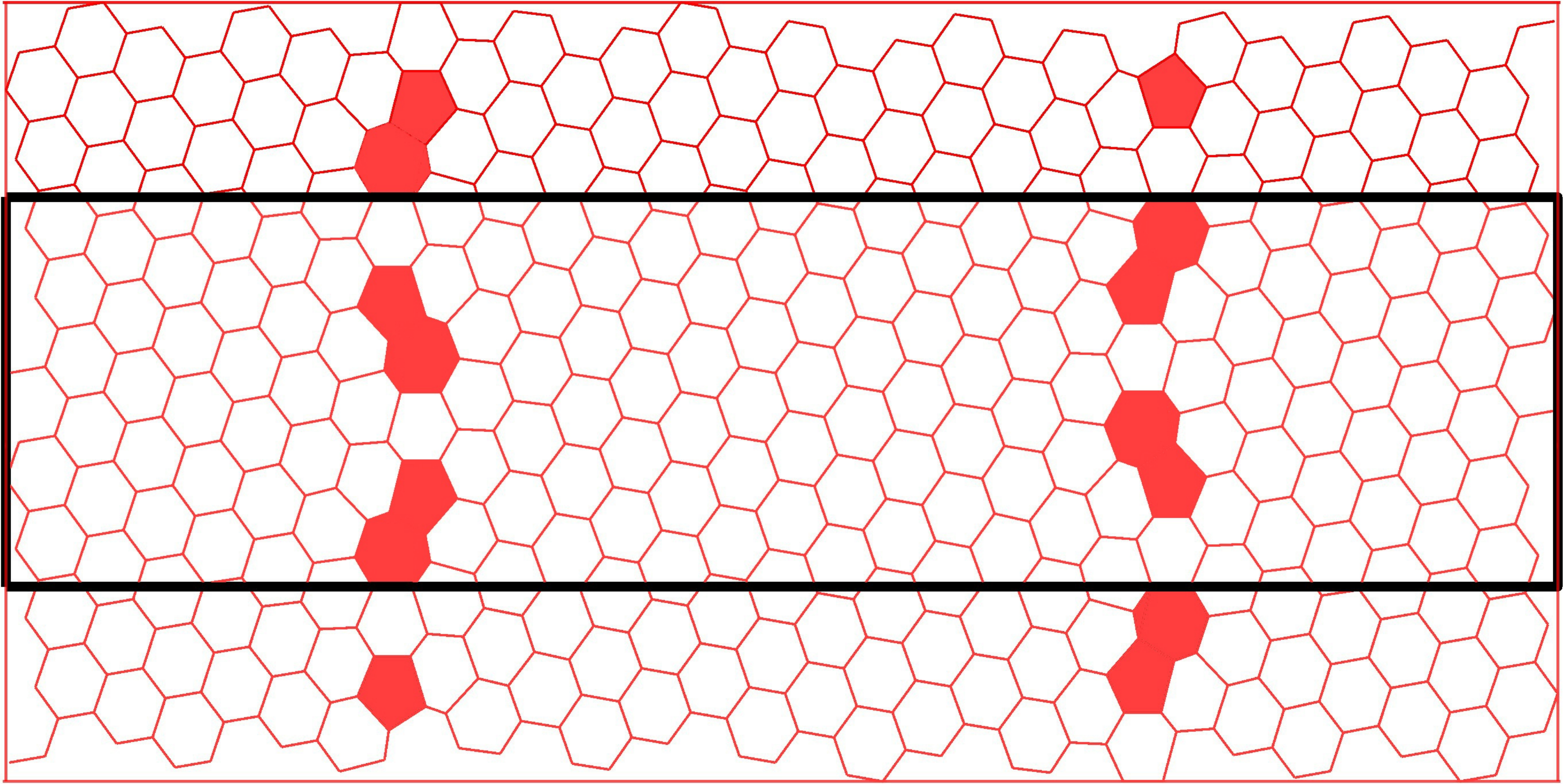}\label{fig01g}}
    \hspace{0.1cm}
    \subfloat[(1,6)~44.82\degree]{\includegraphics[width=0.32\textwidth]{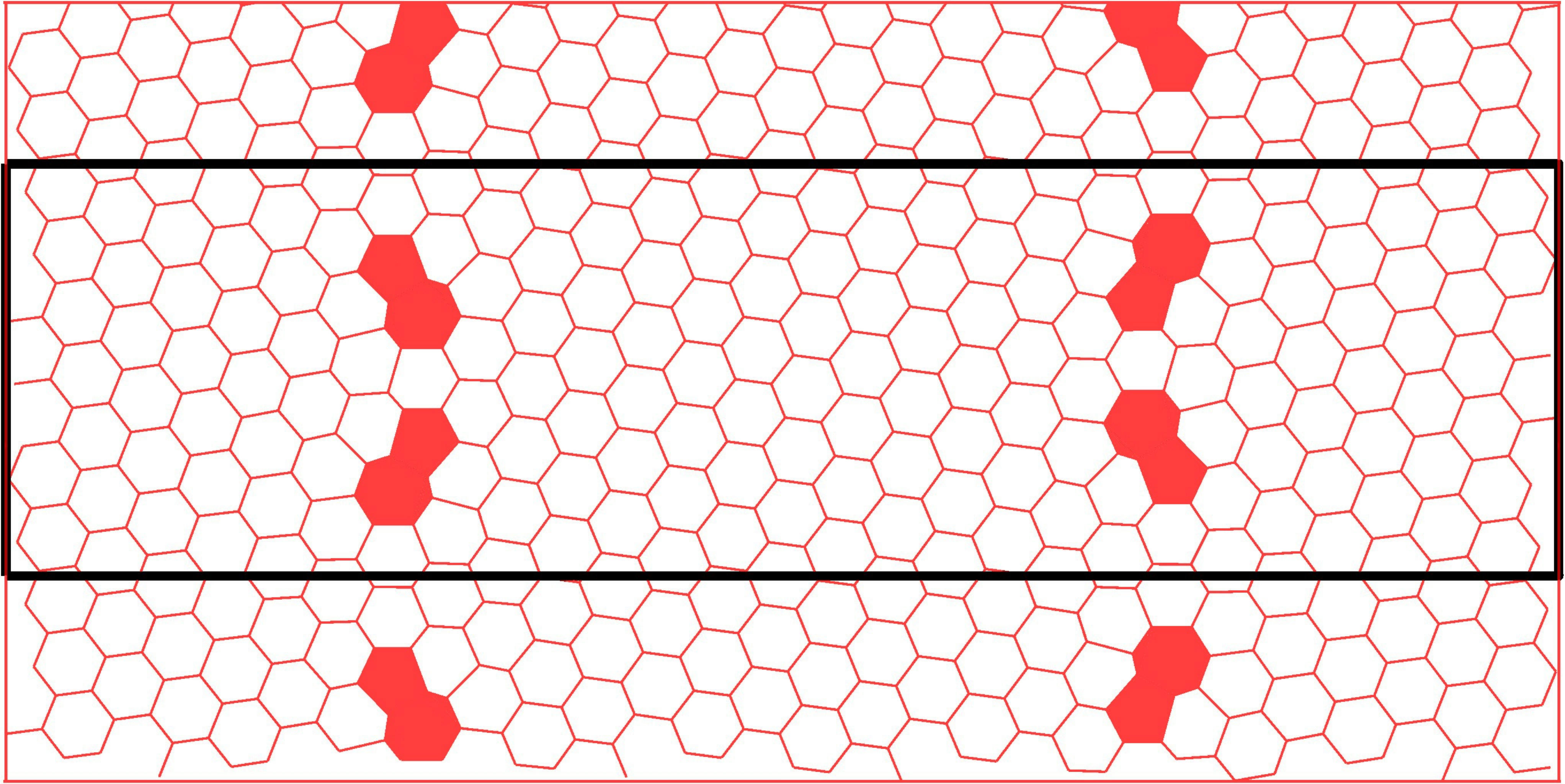}\label{fig01h}}   
    \caption{ Close to $180 \times 32$~\AA$^2$ size  pristine graphene sample and xPC-G samples with seven different misorienation angles. The schematic diagram of the primitive unit 
    cell vectors of the SLG sample is shown in (a) and rectangles created by black boundaries in
    (b)--(h) show the primitive superunit cells of commensurate unit cell for xPC-G samples.} 
    \label{fig01}
\end{figure*}
We follow the centroidal Voronoi tessellation based algorithm proposed by Ophus et al~\cite{ophus2015large} to generate symmetric GBs in graphene.
We prepare xPC-G samples with seven different misorientation angles, where each 
misorientation angle $\theta = \theta_L + \theta_R$ can be defined 
through translation vectors $(n_L, m_L)$ and  $(n_R, m_R)$ corresponding to the 
left and the right domains along the GB, respectively\cite{zhang2013structures}. 
For symmetric GBs, we have $n_L=n_R=n$, $m_L=m_R=m$, 
$\theta_L = \theta_R = \frac{\pi}{3} - \arctan\frac{(2n + m)}{\sqrt{3}}$ and $n$ and $m$ are integers~\cite{zhang2013structures,ophus2015large}.
The corresponding periodic length along the GB direction is 
$L_{y} =a\sqrt{3(n^2+nm+m^2)}$, where the average C-C bond length $a =  1.397$~\AA.
Following Yazyev and Louie~\cite{yazyev2010topological}, the simulation supercell 
for the SED method, involving two parallel equally spaced GBs, is rectangular of 
length $L_x$ and width $L_y$ in order to satisfy periodic boundary conditions (PBCs). 
This means the 
unit cell vectors, $\bold{V}_1 = L_x \hat{\v{i}}$ and $\bold{V}_2 = L_y \hat{\v{j}}$,
of the supercell are orthogonal to each other, 
where $\hat{\v{i}}$ and $\hat{\v{j}}$ are unit vectors in the Cartesian plane. 
Although close to $L_x = 4$~nm is sufficient to maintain stable GB structures, since this
significantly reduces the influence of elastic interactions due to neighbouring GBs ~\cite{blase2000structure, yazyev2010topological}, we have taken 
$L_x = 6$~nm for all our samples. Thus, two neighbouring GBs are separated by 3 nm, which also 
becomes the grain size along $\rm{x}$-axis. The superunit cells for 
$(n,m)$ xPC-G samples with different misorientation angles
$\theta$ have been shown in \figs{fig01b}--\figx{fig01h}. 
The primitive unit cell of the pristine graphene $(1,1)$ sample is created by two planar
lattice vectors  $\bold{V}_1 = \sqrt{3}a \hat{\v{i}}$ and 
$\bold{V}_2 =  \sqrt{3}a(\cos 60^\circ \hat{\v{i}} + \sin 60^\circ \hat{\v{j}})$, as
shown in \fig{fig01a}.

\section{Computational details}
Now, we reiterate some basic principles of the SED method which has been explained in details
in Abhikeern and Singh~\cite{abhikeern2023consistent}.
Based on the Fourier's law and the phonon Boltzmann transport equation (BTE) under relaxation time approximation,  the TC tensor of a system is given by
 \begin{align}
     \v{k} = \sum_{\substack{\gv{\kappa}}} \sum_{\substack{\nu}} c_{v}\mode \left(\v{v}_g\mode \otimes \v{v}_g\mode\right)\tau\mode.
     \label{kSED}
 \end{align}
 Here, the summation is over all allowed normal modes in the first Brillouin zone (BZ), 
 and each mode $\mode$ is denoted by wavevector $\gv{\kappa}$ and  dispersion branch $\nu$. Moreover, $c_{v}\mode$,  $\v{v}_g\mode$ and $\tau\mode$ are the mode specific volumetric specific heat, group velocity and phonon lifetime, respectively. We now explain how we calculate these three physical quantities.  We take $c_{v}\mode 
 =k_B/V$~\cite{mcgaughey2014predicting} for all modes for classical MD simulations, where $k_B$ 
 is the Boltzmann's constant and $V$ is the volume of the simulation box.
 The group velocity $\v{v}_g\mode =\pd{\omega\mode}{\gv{\kappa}}$ is obtained by using finite
 difference method over a fine grid  of wavevectors around the given mode $\mode$. The  GULP
 package~\cite{GULP} is used for calculating the normal mode frequencies $\omega\mode$ and
 the corresponding eigenvectors.
 The phonon lifetime for a mode is calculated by fitting a Lorentzian function over
 the SED curve, which is determined by the normal mode decomposition method by projecting the
 equilibrium MD simulation based atomic positions and velocities onto the normal mode
 coordinates~\cite{abhikeern2023consistent}. 

We describe the simulation details below. 
The primitive superunit cell for $(n,m)$ structure is repeated $N_{\rm{x}}$ and $N_{\rm{y}}$ times along $\rm{x}$ and $\rm{y}$ axes, respectively, to construct a sample of the required size. To ensure the lowest
possible energy state for the GB, we first perform energy minimization over this structure with
PBCs along all axes, where the C-C interactions are modeled by the original REBO potential~\cite{brenner2002second}. 
The equilibrium MD simulation is then performed with multiple MD runs where MD time step was
taken as $\deltmd=0.2$~fs. In the first run, atoms are simulated under NPT ensemble conditions
with zero pressure and 300~K temperature for $2 \times 10^5$ MD steps. The next run is performed
under NVT conditions with $300$~K for $2 \times 10^5$ MD steps. The Nos\'{e}-Hoover chain
thermostats are used for these NPT and NVT runs. The final run is performed for another $2^{16}$
MD time steps under NVE conditions to ensure steady-state conditions without any corrupting
influence of thermostats. The package LAMMPS \cite{lammps_plimpton} is used for all MD simulations and energy minimization. For a given sample, the SEDs are calculated for all
allowable~\cite{abhikeern2023consistent, mcgaughey2014predicting} normal modes in the first BZ.  
Five different MD simulations with different initial atomic velocities are considered so that 
an average of SEDs for all simulations can  be taken for the extraction of the phonon lifetime for a given mode.
Then, we consider only the first quadrant of the hexagonal BZ for SLG and rectangular BZ for xPC-G samples to exploit both the symmetry of the BZ and the simplicity in the discretization of the BZ~\cite{qiu2012molecular,abhikeern2023consistent}. The SEDs of the allowable modes in the 
first quadrant of the first BZ and its symmetric copies in other quadrants are finally averaged over.

\section{Results and Discussion}
Using the primitive superunit cells defined above, we prepare same-size pristine graphene and
seven xPC-G samples with different $\theta$ corresponding to $(n,m)$ translation vectors.
In order to achieve an approximate size of $180 \times 32$~\AA$^2$, the number of unit cells
required along $\rm{x}$ and $\rm{y}$ axes, $N_{\rm{x}}$ and $N_{\rm{y}}$, respectively, for different $(n,m)$ samples
has been listed in Table~\ref{table1}. 
\begin{table}[htp]
\caption{Number of unit cells required along $\rm{x}$ and $\rm{y}$ axes, $N_{\rm{x}}$ and $N_{\rm{y}}$, respectively, to obtain almost $180 \times 32$~\AA$^2$ size samples with different misorientation angles $\theta$.}
\centering
\begin{tabular}{ P{0.085\textwidth}  P{00.085\textwidth}  P{00.085\textwidth}  P{00.2\textwidth}  P{00.085\textwidth}}
\hline
$\theta$ & $(n,m)$  &  $L_y$~(\AA) & Unit cell atoms  & $(\rm{N_x},\rm{N_y})$\\
\hline
0 & (1,1)  & 2.10 & 2 & (73,14) \\
9.43 & (3,4)  & 14.71 & 344 & (3,2) \\
13.17 & (2,3)  & 10.55 & 248 & (3,3) \\
21.78 & (1,2)  & 6.40 & 152 & (3,5)\\
27.80 & (2,5)  & 15.11 & 356 & (3,2) \\
32.20 & (1,3)  & 8.72 & 204 & (3,4) \\
42.10 & (1,5)  & 13.47 & 320 & (3,2) \\
44.82 & (1,6)  & 15.86 & 372 & (3,2) \\
\hline
\end{tabular}
\label{table1}
\end{table}
The table also mentions the unit cell length $L_y$ along 
$\rm{y}$-axis and the number of atoms per unit cell for each sample.

\begin{figure*}[htp]
      \centering
     \subfloat[SLG with 2 basis atoms]{\includegraphics[width=6.5cm, height=5.5cm]{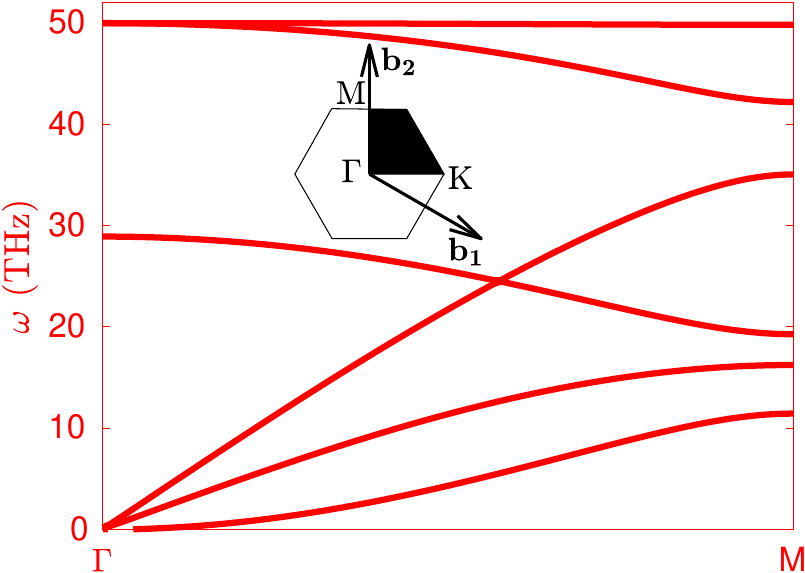}\label{fig02a}}   
     \hspace{0.1cm}
     \subfloat[$21.78^\circ$ ]{\includegraphics[width=0.45\textwidth]{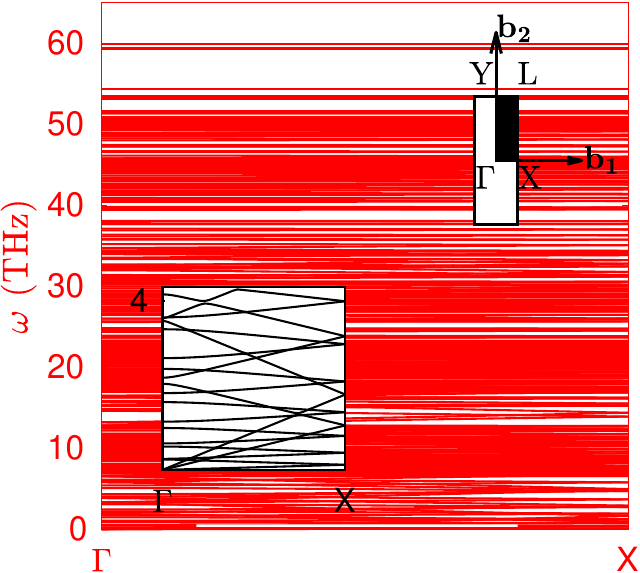}\label{fig02b}}
     \hspace{0.1cm}
     \subfloat[$32.20^\circ$]{\includegraphics[width=0.45\textwidth]{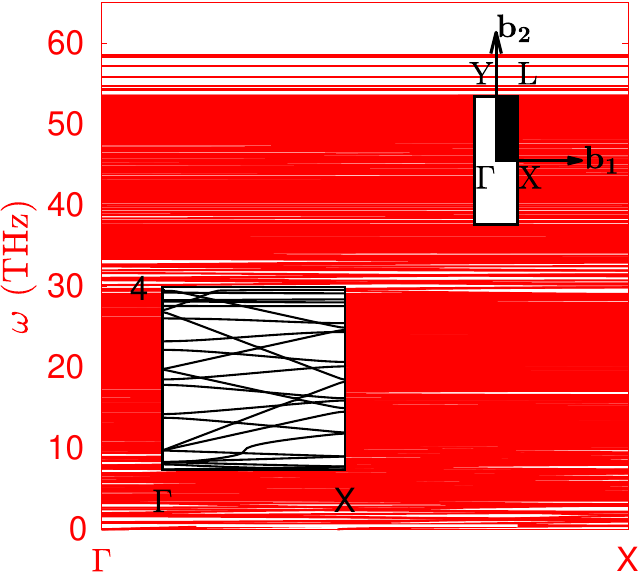}\label{fig02c}} 
     \caption{Dispersion curves in the first BZ for (a) SLG with 2 basis atoms along $\Gamma-\rm{M}$ direction, (b) $21.78^\circ$ xPC-G sample with 152 basis atoms and (c) $32.20^\circ$ xPC-G sample with 204 basis atoms along $\Gamma-\rm{X}$ direction.
     Insets in (b) and (c) show the zoomed-in area of the same dispersion curves with upper limit of frequency constrained to 4~THz. Another set of insets shows the respective BZs
     where the shaded area refers to the first quadrant.}
     \label{fig02}
 \end{figure*}

Based on their respective primitive unit and superunit cells, we first show the phonon dispersion curves in \fig{fig02} for the pristine $(1,1)$ graphene and $(1,2)$ and $(1,3)$ xPC-G samples with $21.78^\circ$ and  $32.20^\circ$ misorientation angles.
The dense dispersion curves in \figs{fig02b} and \figx{fig02c} reflect a large number of 
unit cell atoms, 152 and 204, respectively, for $21.78^\circ$ and $32.20^\circ$ xPC-G samples.
In order to improve clarity, insets in \figs{fig02b} and \figx{fig02c} show the zoomed-in area of the same dispersion curves with upper limit of frequency constrained to 4~THz. 
Similar to Diery et al~\cite{diery2018nature}, we also obtain some high-frequency flat phonon branches.
We also show the respective BZs in another set of insets in \figs{fig02a}--\figx{fig02c}, where 
the shaded area refers to the first quadrant used for discretization of the BZs so that symmetry
of the BZs can be exploited~\cite{qiu2012molecular,abhikeern2023consistent}.

\begin{figure*}[htp]
     \centering
     \subfloat[$21.78^\circ$]{\includegraphics[width=0.4\textwidth]{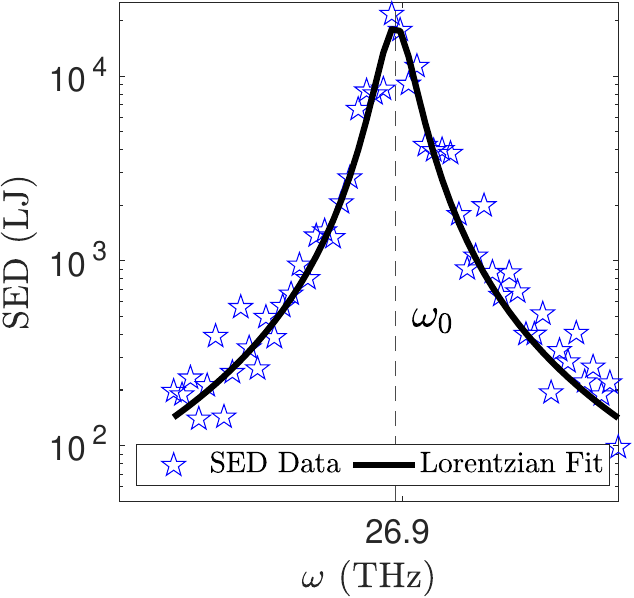}\label{fig03a}}
     \hspace{0.1cm}
     \subfloat[$32.20^\circ$]{\includegraphics[width=0.4\textwidth]{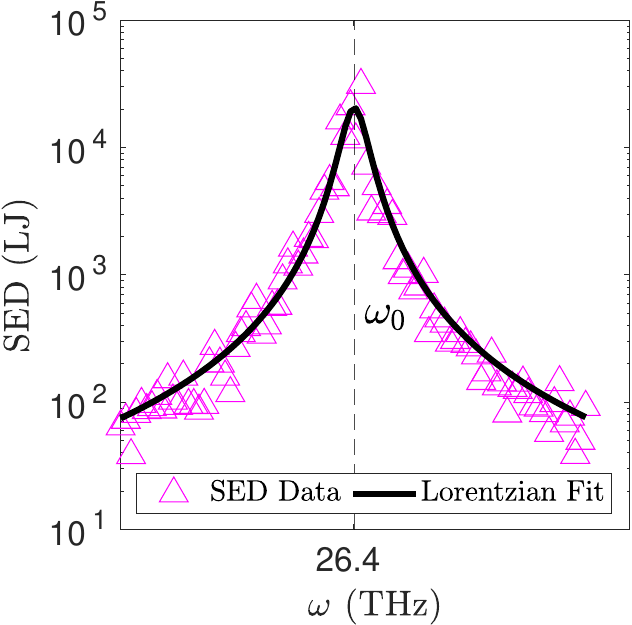}\label{fig03b}}\\
     \hspace{0.1cm}
     \subfloat[$21.78^\circ$]{\includegraphics[width=0.4\textwidth]{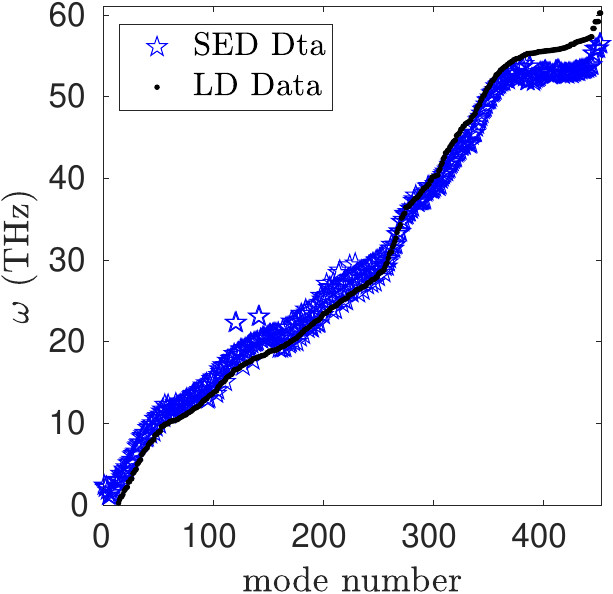}\label{fig03c}}
     \hspace{0.1cm}
     \subfloat[$32.20^\circ$]{\includegraphics[width=0.4\textwidth]{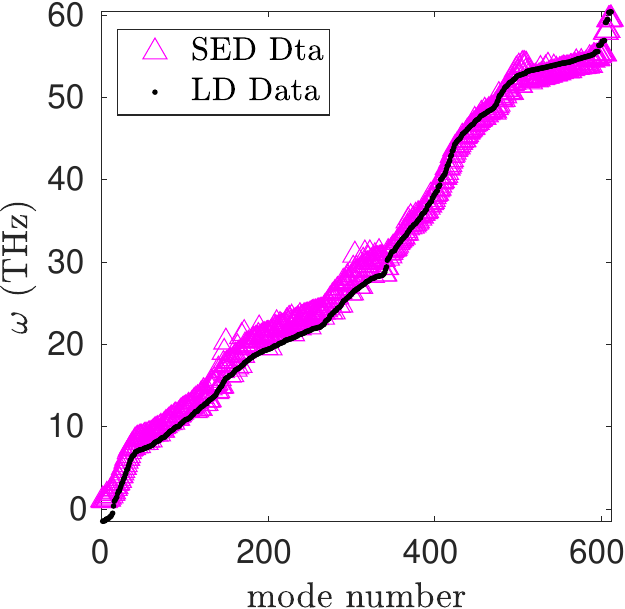}\label{fig03d}}
     \caption{The Lorentzian fit for the SED data obtained at wavevector  $\gv{\kappa}=[\pi/6a,0,0]$ and an optical dispersion branch for (a) $21.78^\circ$ and 
     (b) $32.20^\circ$ xPC-G samples. Comparison between the GULP generated 
     harmonic lattice dynamics frequencies (LD Data) $\omega$ and the Lorentzian 
     curve fitted anharmonic frequencies $\omega_0$ for all available modes
     at X symmetric point with wavevector  $\gv{\kappa}=[\pi/a,0,0]$.
     }
     \label{fig03}
 \end{figure*}

For all our samples, we obtain excellent Lorentzian fits for the SED curves. 
We particularly show the Lorentzian curve fitting over the SED data in \fig{fig03} for $21.78^\circ$ and $32.20^\circ$ xPC-G samples for a chosen wavevector 
$\gv{\kappa}=[\pi/6a,0,0]$ and one of the optical dispersion branches.
The fitting provides not only the phonon lifetime $\tau$ but also the 
anharmonic frequency $\omega_0$ as the center of the Lorentzian fit over the SED 
curve. These anharmonic frequencies match closely with the harmonic lattice dynamics (GULP)
based frequencies. For illustration, we provide excellent match for all available modes 
for a given wavevector $\gv{\kappa}$ around the ``X'' point as shown in \figs{fig03c}
and \figs{fig03d} for  $21.78^\circ$ and $32.20^\circ$ angles, respectively. Hence,
it is established that the harmonic frequencies are a good approximation for the
vibrational modes at the room temperature.

\begin{figure}[htp]
    \centering     
    \subfloat[$\tau$]{\includegraphics[width=0.45\textwidth]{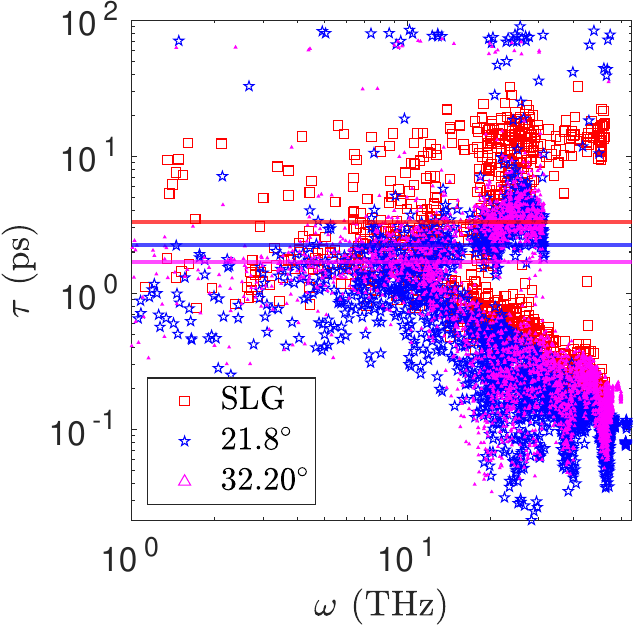}\label{fig04a}} 
    \hspace{0.1cm}    
    \subfloat[$\rm{v}_g$]{\includegraphics[width=0.435\textwidth]{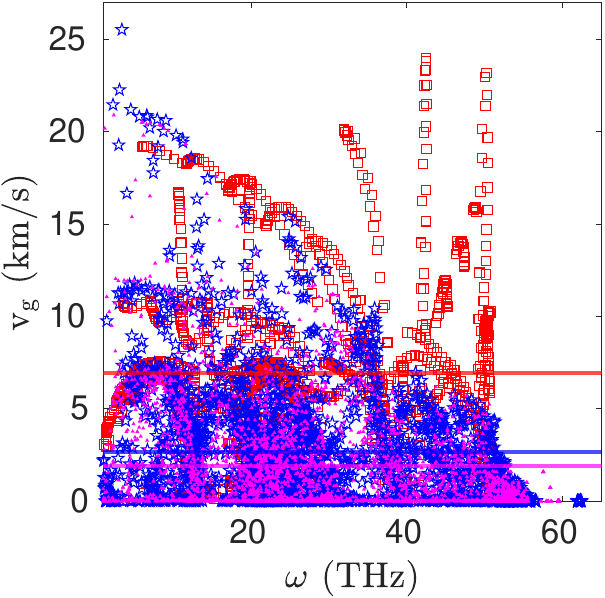}\label{fig04b}} 
    \hspace{0.1cm}
    \subfloat[$\rm{l_x}$]{\includegraphics[width=0.45\textwidth]{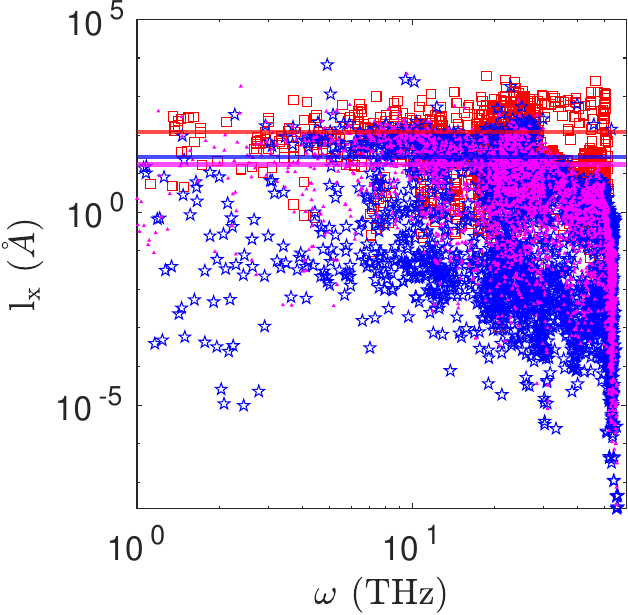}\label{fig04c}}
    \hspace{0.1cm}
    \subfloat[$\rm{l_y}$]{\includegraphics[width=0.45\textwidth]{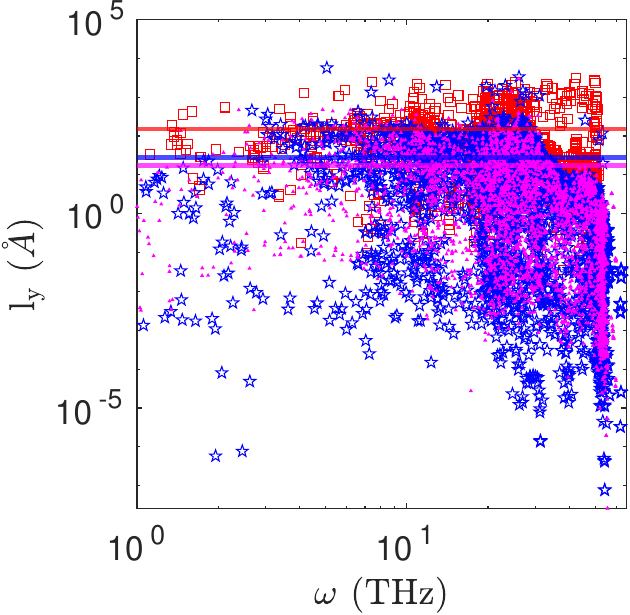}\label{fig04d}}
    \caption{(a) Phonon lifetimes in ps, (b) group velocities in km/s, and mean free paths
    (c) $\rm{l_x}$ and (d) $\rm{l_y}$ of the allowable normal modes in the 
    the first quadrant of the BZs for pristine SLG (red) and xPC-G  with $\theta = 21.78^\circ$ (blue) and $\theta= 32.20^\circ$ (magenta) samples, which are prepared with $N_{\rm{x}} \times N_{\rm{y}} $ primitive unit cells, 
    where $N_{\rm{x}} = 73, 3, 3$ and $N_{\rm{y}} = 14, 5, 4$, 
    respectively, for three systems. The solid horizontal lines show the average values. }
    \label{fig04}
\end{figure}

\begin{figure}[htp]
    \centering     
    \subfloat[$\rm{k}_{xx}$ and $\rm{k}_{yy}$]{\includegraphics[width=5cm, height=5cm]{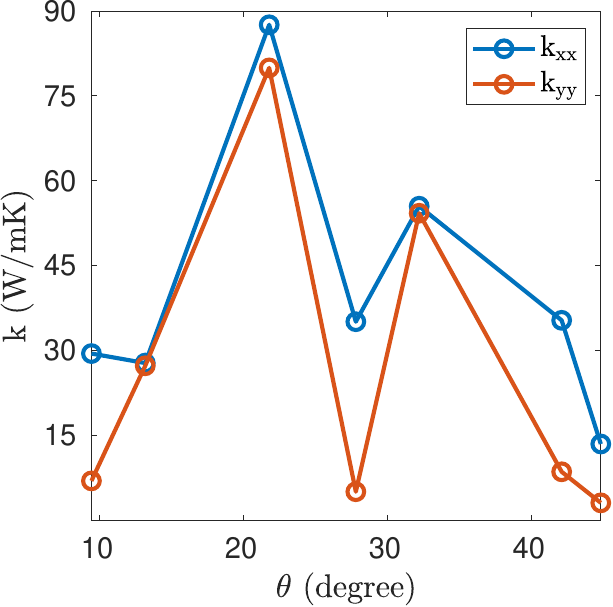}\label{fig05a}} 
    \hspace{0.1cm}
    \subfloat[$\rm{v}_{g_{\rm{rms}}}$ along x and y axes]{\includegraphics[width=5cm, height=5cm]{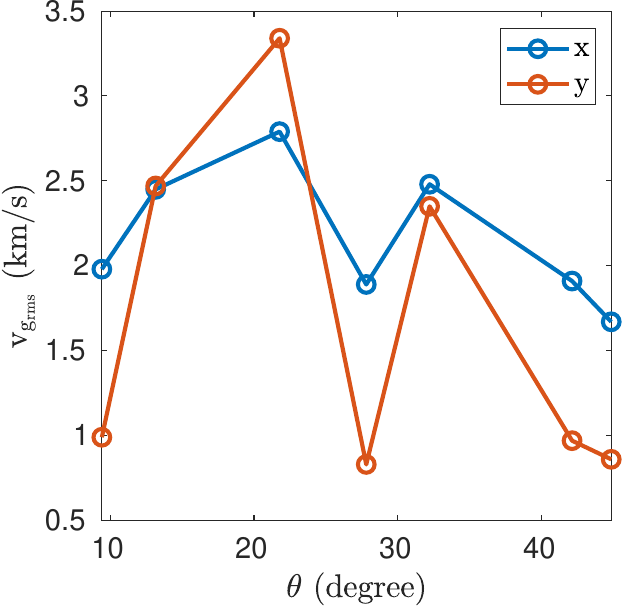}\label{fig05b}} \\
    \subfloat[$\avg{\tau}$]{\includegraphics[width=5cm, height=5cm]{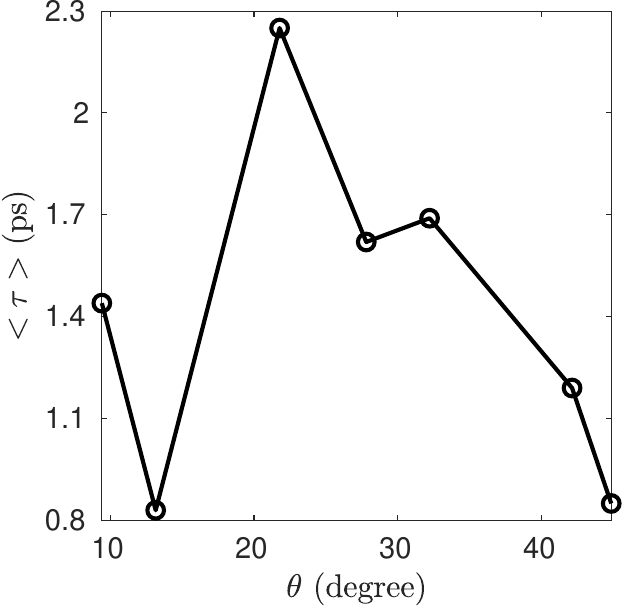}\label{fig05c}}
    \hspace{0.1cm}
    \subfloat[$\rho$]{\includegraphics[width=5cm, height=5cm]{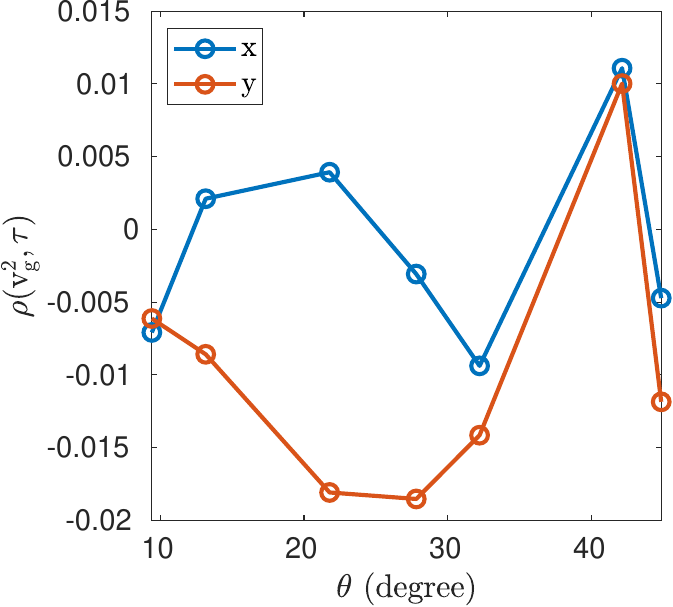}\label{fig05d}}\\
    \subfloat[$\rm{k}_{\rm{I}}$ and $\rm{k}_{\rm{II}}$ components]{\includegraphics[width=5.4cm, height=5cm]{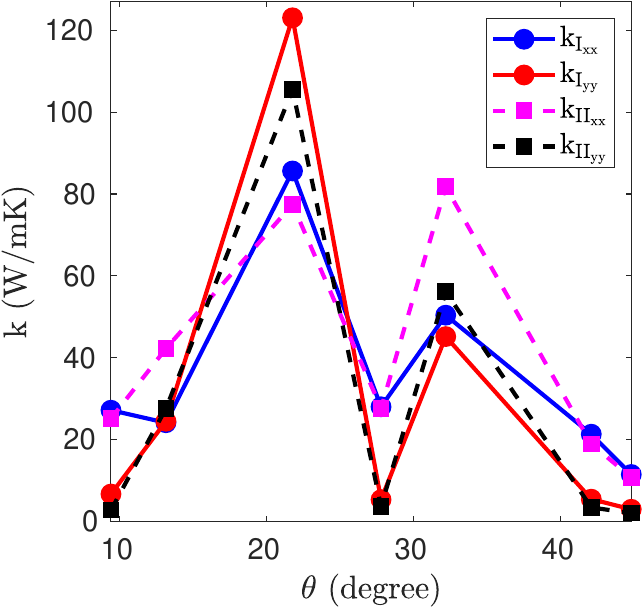}\label{fig05e}}
    \hspace{0.1cm}
    \caption{ For all misorientation angles, (a) $\rm{k}_{xx}$ and $\rm{k}_{yy}$, (b) x and y components of $\v{v}_{g_{\rm{rms}}}$, (c) $\avg{\tau}$, (d) correlation coefficients $\rho(\rm{v}_{g_x}^2, \tau)$ and
    $\rho(\rm{v}_{g_y}^2, \tau)$ , (e) $\rm{k}_{\rm{I}_{xx}}$ and $\rm{k}_{\rm{I}_{yy}}$ in solid lines and $\rm{k}_{\rm{II}_{xx}}$ and $\rm{k}_{\rm{II}_{yy}}$ in dashed lines.}
    \label{fig05}
\end{figure}

The phonon lifetimes ($\tau$), group velocities $\rm{v}_g = \left( \sqrt{\rm{v}_{g_x}^2 + \rm{v}_{g_y}^2} \right)$ and mean free paths along $\rm{x}$ and $\rm{y}$ axes, $\rm{l_x} = \rm{v}_{g_x}\tau$ and $\rm{l_y}=\rm{v}_{g_y}\tau$, obtained for the allowable modes
in the first quadrant of the first BZs of pristine graphene and two xPC-G samples with 
$\theta = 21.78^\circ$ and  $32.20^\circ$ have been shown in \fig{fig04}.
With a total of $1533, 1710$ and $1836$ modes for these three systems, respectively, it makes
more sense to talk about root mean square (rms) of group velocities $\rm{v}_{g_{\rm{rms}}}$,
average of phonon lifetimes $\avg{\tau}$ and average of mean free paths, $\avg{\rm{l_x}}$ and
$\avg{\rm{l_y}}$, rather than individual acoustic or optical modes, where the average is taken
over all available modes in the first quadrant of the first BZ.
We obtain $\avg{\tau}=3.33,2.25,1.69$~ps, $\rm{v}_{g_{\rm{rms}}} = 6.93,2.67,1.91$~km/s, $\avg{\rm{l_x}}=121.1,27.7,17.3$~\AA~ and $\avg{\rm{l_y}}=152.1,28.0,17.5$~\AA, respectively,
for three systems shown in \fig{fig05} with solid horizontal lines. 
Later, we also show $\rm{x}$ and $\rm{y}$ components of $\v{v}_{g_{\rm{rms}}}$ in \fig{fig05b} and $\avg{\tau}$ in \fig{fig05c}for all angles. 
We note that average mean free paths are less than the grain size along $\rm{x}$-axis (30~\AA) for xPC-G samples. We also observe that around 90\% of
the phonon modes have their mean free paths less than 30~\AA~ along both $\rm{x}$ and $\rm{y}$ axes for
all seven different misorientation angles. This was expected for $\rm{l}_x$ since phonons would
scatter at the GBs, however, $\rm{l}_y$ also showing similar trend is an interesting result.

Having obtained all phonon properties, we use \eqn{kSED} to obtain the TC tensor components $\rm{k_{xx}}\approx 87, 29, 28, 88, 35,56,25, 14$~W/mK and $\rm{k_{yy}} \approx 80, 
7, 27, 80, 5, 54$, $9, 3$~W/mK for pristine graphene and seven different misorientation angles 
listed in Table~\ref{table1} in increasing order. With respect to the pristine graphene,
maximum reduction in the TCs is found in the last sample with $44.82^\circ$ angle, almost $84\%$ and
$96\%$ reduction for $\rm{k}_{xx}$ and $\rm{k}_{yy}$, respectively. 
As reflected in the values and also shown in \fig{fig05a}, the TCs strongly depend upon the  
misorientation angles. 
We also note that except for samples with $13.17^\circ$ and $32.2^\circ$ tilt angles, 
there exists strong anisotropy along $\rm{x}$ and $\rm{y}$ axes in the TCs for other angles. 
Also, the TCs for $21.78^\circ$ sample is the highest among all xPC-G samples. These trends
in TCs can be explained in terms of the trends in the group velocities and the phonon lifetimes, which were  found to be uncorrelated for all xPC-G samples. Since 
$\rm{k}_{xx} = \sum_{\substack{\gv{\kappa},\nu}} c_{v}\mode \rm{v}_{g_x}^2\mode\tau\mode$, and we have taken $c_{v}\mode = k_B/V=c_v$ for all modes, hence, the average of $\rm{k}_{xx}$ over all $n$ modes in the firxt BZ can be written as $\avg{\rm{k}_{xx}}= n c_v \avg{\rm{v}_{g_x}^2} \avg{\tau}= n c_v\rm{v}_{g_{x_{\rm{rms}}}}^2 \avg{\tau}$, provided that correlation coefficient $\rho$ of $\rm{v}_{g_x}^2$ and $\tau$ for all modes is close to zero. The similar
argument holds for $\avg{\rm{k}_{yy}}$. The averages of TCs defined in this way are denoted as
$\avg{\rm{k}_{xx}}_{\rm{I}}$ and $\avg{\rm{k}_{yy}}_{\rm{I}}$.
We calculate $\rho(\rm{v}_{g_x}^2, \tau)$ and $\rho(\rm{v}_{g_y}^2, \tau)$ and plot them for all angles in \fig{fig05d}. All correlation coefficients were found to be in the range of $0.2\%-1.9\%$, i.e. they are close to zero. This suggests that group velocities and phonon
lifetimes are uncorrelated for all xPC-G samples. 
We obtain  $\avg{\rm{k}_{xx}}_{\rm{I}} \approx 27, 24, 86, 28,50,21, 12$~W/mK and $\avg{\rm{k}_{yy}}_{\rm{I}} \approx  7, 25, 123, 5, 45, 6, 3$~W/mK for seven different angles in
the increasing order. They are also plotted  in \fig{fig05e} and we observe that values closely match with actual $\rm{k}_{xx}$ and $\rm{k}_{yy}$ reported earlier, except for $21.78^\circ$
sample for which $\avg{\rm{k}_{yy}}_{\rm{I}} = 123$~W/mK whereas $\rm{k}_{yy} = 80$~W/mK,
i.e. an increase of around 54\% is found. This deviation can be attributed to $21.78^\circ$ sample having a slightly higher $\rho(\rm{v}_{g_y}^2, \tau)$ and the highest $\rm{v}_{g_{y_{\rm{rms}}}}$. 
But in general, we can assert that $\avg{\rm{k}_{xx,yy}}=\avg{\rm{k}_{xx,yy}}_{\rm{I}} \approx \rm{k}_{xx,yy}$ and  it can be emphasized that an average phonon mode can be defined whose group
velocity components along $\rm{x}$ and $\rm{y}$ axes are the rms of group velocity components of all
phonon modes and whose lifetime is the average of all phonon lifetimes. Therefore, anisotropy in
the TC components is largely the outcome of anisotropy in the group velocities, since lifetimes
remain the same for both $xx$ and $yy$ components of the TC. The maxima and minima of the TC
components across all angles, however, have to explained both in terms of maxima and minima of
$\rm{x}$ and $\rm{y}$ components of $\v{v}_{g_{\rm{rms}}}$ and also those of $\avg{\tau}$. 

\begin{figure}[htp]
    \centering     
    \subfloat[DOS]{\includegraphics[width=5.95cm, height=5.95cm]{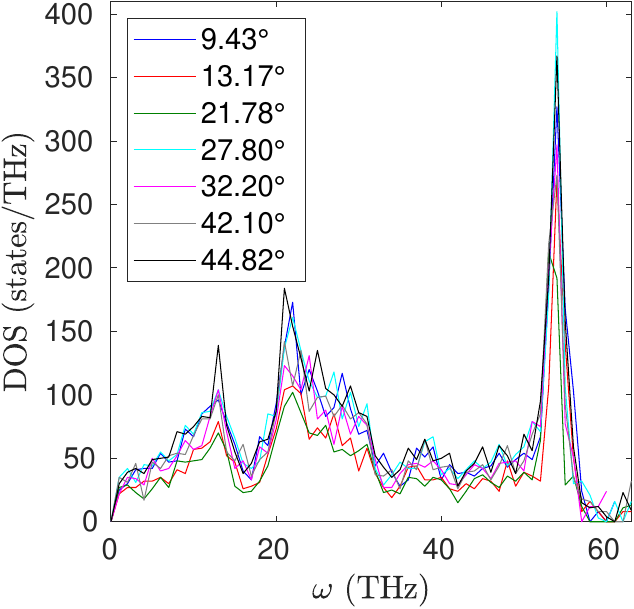}\label{fig06a}} 
    \hspace{0.1cm}
    \subfloat[$\overline{\rm{v}_{g_{x_{\rm{rms}}}}}$]{\includegraphics[width=6cm, height=6cm]{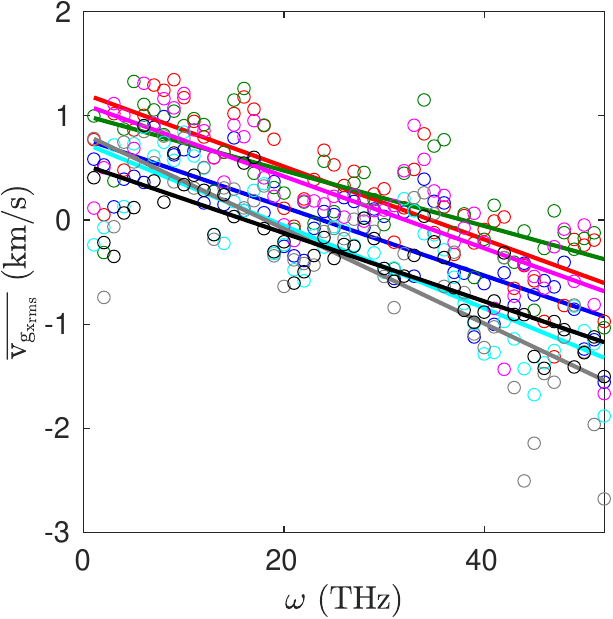}\label{fig06b}}\\
    \subfloat[$\overline{\rm{v}_{g_{y_{\rm{rms}}}}}$]{\includegraphics[width=6cm, height=6cm]{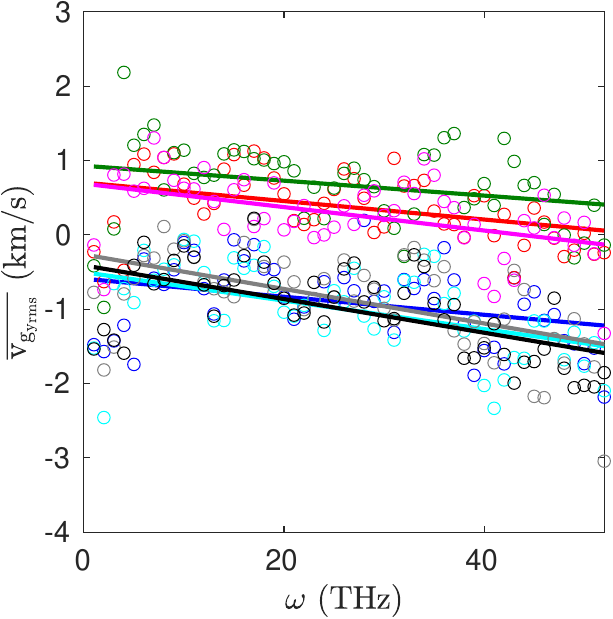}\label{fig06c}}
    \hspace{0.1cm}
        \subfloat[$\overline{\tau}$]{\includegraphics[width=6cm, height=6cm]{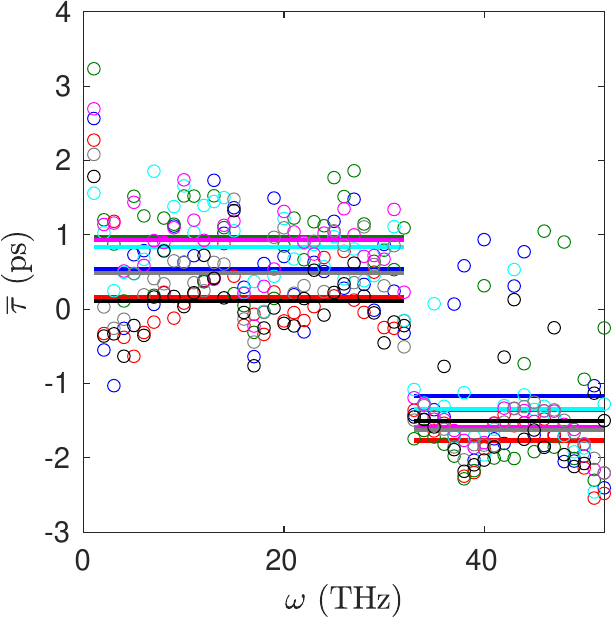}\label{fig06d}} 
    \caption{ (a) DOS, (b) $\overline{\rm{v}_{g_{x_{\rm{rms}}}}}$, (c) $\overline{\rm{v}_{g_{y_{\rm{rms}}}}}$, (d) $\overline{\tau}$ for all
    xPC-G samples of size close to $180 \times 32$~\AA$^2$.}
    \label{fig06}
\end{figure}

Next, we express group velocity and phonon lifetime as distribution functions of normal mode frequencies so
that the TC components can also be written as distribution functions of frequencies. This can potentially
help in the semi-analytical and numerical solutions of the BTE~\cite{hua:2015}. For that, we first show
phonon density of states (DOS) in \fig{fig06a}, where $\rm{DOS}(\omega)d\omega$ represents the number of 
modes lying between $\omega$ and $\omega+d\omega$.
We observe that the DOS remains almost the same for all GB angles~\cite{cao2012kapitza}. 
This is followed by calculating the sum of $\v{v}_{g_{\rm{rms}}}$ along $\rm{x}$ axis for 
all modes  lying in the frequency range $[\omega,\omega+ d\omega)$ and then dividing 
by the total number of modes in the same range, which we denote as 
$\overline{\rm{v}_{g_{x_{\rm{rms}}}}}$. Similarly, we define 
$\overline{\rm{v}_{g_{y_{\rm{rms}}}}}$ and $\overline{\tau}$  for $\rm{v}_{g_{\rm{rms}}}$ along $\rm{y}$ axis and $\tau$, respectively. Basically, they are distribution functions of $\omega$ and 
can be treated as average properties of phonons in  $[\omega,\omega+ d\omega)$. 
For any given angle, we find that natural semi-log plots of 
$\overline{\rm{v}_{g_{x_{\rm{rms}}}}}(\omega)$ and $\overline{\rm{v}_{g_{y_{\rm{rms}}}}}(\omega)$
can be fitted with straight lines as shown in \figs{fig06b} and \figx{fig06c}, respectively,
and the natural semi-log plot of $\overline{\tau}(\omega)$ can be fitted with a piecewise
constant function in \fig{fig06d}. Therefore, we can write, 
\begin{subequations}
\begin{alignat}{3}
\overline{\rm{v}_{g_{x_{\rm{rms}}}}} (\omega)&= a_1e^{-a_2\omega}, \quad \omega \in [0,\omega_{\rm{max}}]\\
\overline{\rm{v}_{g_{y_{\rm{rms}}}}}(\omega) &= b_1e^{-b_2\omega}, \quad \omega \in [0,\omega_{\rm{max}}]\\
\overline{\tau}(\omega) &= 
\begin{cases}
    c_1, \quad \omega \in [0,\omega_d]\\
    c_2, \quad \omega \in (\omega_d,\omega_{\rm{max}}]\\
\end{cases}
\end{alignat}
\label{eq2}
\end{subequations}
where  $\omega$'s are in THz, group velocity components are in km/s, lifetimes are in ps, 
$a_1, a_2, b_1, b_2, c_1, c_2$ are constants,  $\omega_d < \omega_{\rm{max}}$ is the frequency at which discontinuity in $\overline{\tau}$ emerges and $\omega_{\rm{max}}$ is the
maximum frequency beyond which the phonon properties can be assumed as zero. Interestingly, we find that $\omega_d = 32$~THz and $\omega_{\rm{max}}=52$~THz are the same for all xPC-G samples. The values of the constants appearing in \eqn{eq2} for different misorientation 
angles are provided in Table~\ref{table2}.
\begin{table}[htp]
\caption{$a_1$, $a_2$, $b_1$, $b_2$, $c_1$, $c_2$ for different GB angles.}
\centering
\begin{tabular}{ P{0.060\textwidth}  P{0.060\textwidth}  P{00.060\textwidth} P{00.060\textwidth}  P{00.060\textwidth}  P{00.060\textwidth}  P{00.060\textwidth}  }
\hline
$\theta$ &$a_1$ & $a_2$ & $b_1$ & $b_2$ & $c_1$ & $c_2$ \\
\hline
9.43 & 2.182 & 0.033 & 0.554 & 0.012 & 1.704 & 0.311 \\
13.17 & 3.360 & 0.035 & 2.020 & 0.012 & 1.170 & 0.171 \\
21.78 & 2.735 & 0.027 & 2.534 & 0.010 & 2.654 & 0.259 \\
27.80 & 2.094 & 0.040 & 0.605 & 0.019 & 2.298 & 0.260 \\
32.20 & 3.031 & 0.035 & 1.992 & 0.016 & 2.550 & 0.205 \\
42.10 & 2.282 & 0.045 & 0.768 & 0.023 & 1.624 & 0.197 \\
44.82 & 1.699 & 0.033 & 0.663 & 0.023 & 1.115 & 0.222 \\
\hline
\end{tabular}
\label{table2}
\end{table}
Based on this, we calculate another measure of the TC components as
\begin{align}
\avg{\rm{k}_{xx,yy}}_{\rm{II}} = c_v\int_0^{\omega_{\rm{max}}} \textrm{DOS}(\omega) \left(\overline{\rm{v}_{g_{{x,y}_{\rm{rms}}}}}\right)^2  \overline{\tau}(\omega)d\omega,
\label{eq3}
\end{align}
where $c_v=k_B/V$ as defined above.
The values obtained for xPC-G samples with increasing angles are 
$\avg{\rm{k}_{xx}}_{\rm{II}}\approx 25, 42, 77, 28, 82, 19, 11$~W/mK and
$\avg{\rm{k}_{yy}}_{\rm{II}}\approx 3, 28, 105, 4, 56, 4, 2$~W/mK.
We find that $\avg{\rm{k}_{xx,yy}}_{\rm{II}}$ vary within $\pm 60$\%
range of the actual $k_{xx,yy}$ calculated by \eqn{kSED}, and therefore,
we can assert that $\avg{\rm{k}_{xx,yy}}_{\rm{II}}\approx \rm{k}_{xx,yy} \pm\,0.6~\rm{k}_{xx,yy}$.

\begin{figure}[htp]
\centering
\includegraphics[width=0.6\textwidth]{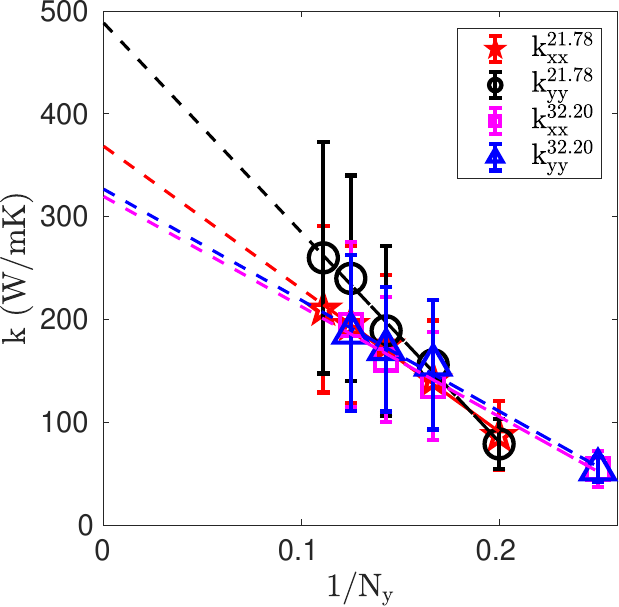}
  \caption{Size dependency of the TCs of the xPC-G samples with  $\theta = 21.78^\circ$ and 
  $\theta = 32.20^\circ$. The dashed lines are showing the extrapolation to y-axis to obtain the bulk TC.}
  \label{fig07}
\end{figure}

The TCs obtained through the SED method are size-dependent~\cite{abhikeern2023consistent}. 
A size-dependent analysis is performed for xPC-G samples with 
$\theta=21.78^\circ$ and $32.20^\circ$ by increasing the number of primitive superunit cells along $\rm{y}$-axis, $N_{\rm{y}}$, but keeping $N_{\rm{x}}$ same because we reason that the $\rm{x}$-axis  length, around 
18~nm, is too large to significantly affect the TC components if we further increase $N_{\rm{x}}$.
In Ref.~\cite{abhikeern2023consistent}, we calculated the TC for SLG up to $50$~\AA~ length 
along $\rm{x}$-axis, and the TC deviated from the bulk value by only 10\%. 
We chose these two angles because they have the least number of basis atoms in comparison 
to xPC-G samples with other misorientation angles, specifically, 152 for
$21.78^\circ$ and 204 for $32.20^\circ$, which makes the computation more tractable.
We take $N_{\rm{y}}=5,6,7,8~\rm{and}~9$ for $21.78^\circ$ samples and $N_{\rm{y}}=4,6,7~\rm{and}~8$ for $32.20^\circ$ samples.
The results for TC components $\rm{k}_{\rm{xx}}$ and $\rm{k}_{\rm{yy}}$ are shown in \fig{fig07}.
Following Ref.~\cite{abhikeern2023consistent}, we perform a linear fit between $\rm{k}_{xx,yy}$ and 
$1/N_{\rm{y}}$ and then extrapolate the linear curve to obtain the bulk TC ( $\rm{k}_{xx,yy}^\infty$).
We calculate  $\rm{k}_{xx}^\infty = 369$ and $320$~W/mK, and $\rm{k}_{yy}^\infty = 488$ and $327$~W/mK for $\theta=21.78^\circ$ and $32.20^\circ$, respectively. 
The bulk TC values for $32.20^\circ$ match closely with Fox et al~\cite{fox2019thermal} who
calculated around 312~W/mK for $32.20^\circ$ BC-G samples with the NEMD method.
For the pristine graphene studied here, 
$\rm{k}_{xx}^\infty = 563$ and $\rm{k}_{yy}^\infty = 863$~W/mK. Therefore, 
$\rm{k}_{xx}^\infty$ decreased by 34\% and 43\% and $\rm{k}_{yy}^\infty$ decreased by 43\% and 62\%, respectively, for the two angles with respect to the pristine graphene. 
We also note that the $32.20^\circ$ sample does not show much anisotropy in the bulk TC 
components whereas the $21.78^\circ$ sample shows strong anisotropy. 

\section{Conclusion}
In conclusion, we predict the phonon properties of xPC-G samples with seven different
misorientation angles with the equilibrium MD simulations and the SED method. 
Our work concludes that the TCs of xPC-G samples strongly depend upon misorientation angles.
We also find that the square of the group velocity components along 
$\rm{x}$ and $\rm{y}$ axes and the phonon lifetimes are uncorrelated for all samples, which 
allows for the calculation of TCs
in terms of an average phonon mode whose group velocity components are the
rms of group velocity components of all phonon modes and lifetime is
the mean of all phonon lifetimes.  
We explain anisotropy in the TC components for some angles with the 
difference in the rms of group velocity components of all phonon modes along $\rm{x}$ and $\rm{y}$ axes. 
Further, the DOS for all 
xPC-G samples are found to be independent of the misorientation angles. Based on the DOS, 
distribution functions of phonon properties have been calculated and
plotted as semi-log plots against the phonon normal frequencies.
The distributions of group velocity components are found to be exponentially decaying whereas the 
distribution of phonon lifetime showed piecewise constant function behavior with respect to frequency. 
This is reflected in the distribution of TC components, which also show exponentially decaying behavior
against frequency.
We provide the parameters for the exponential and piecewise constant functions for the 
group velocity components and lifetimes, respectively.
We provide another measure of the TC based on these functions and estimate that their values are within $\pm 60\%$ range of 
the actual TC values.
Finally, 
we perform the size-dependent analysis for two angles, $21.78^\circ$ and $32.20^\circ$,
and calculated their bulk TC components, which are found to have decreased by 34\% to 62\%
in comparison to the bulk TC values of the pristine graphene. 
We end with a suggestion that the frequency dependent distribution functions suggested
for the rms of group velocity components and the phonon 
lifetime can be used in multiscale semi-analytical and numerical 
solutions for the solutions of the BTE.

\section*{Acknowledgment}
This work was supported by the Department of Science \& Technology, India under the SERB grant MTR/2021/000550. Authors gratefully acknowledge the financial supports and the computing resources of IIT Bombay. Views expressed in this paper are those of the authors and neither of their affiliated Institution nor of the funding agencies.

\bibliographystyle{elsarticle-num}
\bibliography{references}

\end{document}